\documentclass[twocolumn,showpacs,superscriptaddress,preprintnumbers,amsmath,amssymb]{revtex4}
\usepackage{graphicx}% Include figure files
\usepackage{dcolumn}% Align table columns on decimal point
\usepackage{bm}% bold math
\begin{document}
\title{Spin transition in the fractional quantum Hall regime: Effect of extent of the wave function}
\author{V.~V. Vanovsky}
\affiliation{Institute of Solid State Physics, Chernogolovka, Moscow District 142432, Russia}
\affiliation{Moscow Institute of Physics and Technology, Dolgoprudny, Moscow District 141700, Russia}
\author{V.~S. Khrapai}
\affiliation{Institute of Solid State Physics, Chernogolovka, Moscow District 142432, Russia}
\affiliation{Moscow Institute of Physics and Technology, Dolgoprudny, Moscow District 141700, Russia}
\author{A.~A. Shashkin}
\affiliation{Institute of Solid State Physics, Chernogolovka, Moscow District 142432, Russia}
\author{V. Pellegrini}
\affiliation{NEST, Istituto Nanoscienze-CNR and Scuola Normale Superiore, Piazza San Silvestro 12, I-56127 Pisa, Italy}
\author{L. Sorba}
\affiliation{NEST, Istituto Nanoscienze-CNR and Scuola Normale Superiore, Piazza San Silvestro 12, I-56127 Pisa, Italy}
\author{G. Biasiol}
\affiliation{CNR-IOM, Laboratorio TASC, Area Science Park, I-34149 Trieste, Italy}
\begin{abstract}
Using a magnetocapacitance technique, we determine the magnetic field
of the spin transition, $B^*$, at filling factor $\nu=2/3$ in the 2D
electron system in GaAs/AlGaAs heterojunctions. The field $B^*$ is
found to decrease appreciably as the wave function extent controlled
by back gate voltage is increased. Our calculations show that the
contributions to the shift of $B^*$ from the change of the Coulomb
energy and the $g$ factor change due to nonparabolicity are
approximately the same. The observed relative shift of $B^*$ is
described with no fitting parameters.
\end{abstract}
\maketitle

Spin degrees of freedom are important in determining the ground
states and excitations of the fractional quantum Hall effect in not
too strong magnetic fields. In view of the competition between the
Coulomb energy in the two-dimensional (2D) electron system,
$E_C\propto e^2/\kappa l_B$ (where $\kappa$ is the dielectric
constant and $l_B=(\hbar /eB)^{1/2}$ is the magnetic length), and
the Zeeman energy, $E_Z=g\mu_BB$, the fully polarized states in the
high-field limit should become depolarized with decreasing magnetic
field. This manifests itself as spin transitions between fully and
partially polarized states and between states with different partial
polarization. The ratio of the Zeeman and Coulomb energies is varied
in experiments either by changing both the magnetic field, $B$, and
the electron density, $n_s$, at fixed filling factor, $\nu=n_sh/eB$,
or by introducing a parallel component of the magnetic field.
Ground-state spin transitions for the 2D electrons in GaAs have been
observed using activation energy measurements at filling factor
$\nu=2/3$, 4/3, 3/5, etc. \cite{eisenstein90,clark90,engel92};
optical spectroscopy studies at $\nu=2/3$, 2/5, 3/5, etc.
\cite{kukushkin99}; nuclear magnetic resonance measurements at
$\nu=2/3$ \cite{freytag01}; optical absorption experiments at
$\nu=2/3$ \cite{groshaus07}. The spin transitions occur at very
different magnetic fields for different samples. For $\nu=2/3$, the
transition observed in
Refs.~\cite{eisenstein90,clark90,engel92,kukushkin99} is the case at
relatively low fields 2--4~T, while the transition found in
Refs.~\cite{freytag01,groshaus07} occurs in considerably higher
fields, about 8~T or yet higher. In contrast to the $\nu=2/3$ state,
the quantum Hall state at $\nu=1/3$ is found to be fully spin
polarized for all magnetic fields \cite{groshaus07}. Still, spin-flip
excitations at $\nu=1/3$ have been observed in transport
\cite{leadley97}, optical spectroscopy \cite{kukushkin00}, and
inelastic light scattering measurements \cite{groshaus08}.

Numerical calculations based on the composite fermion model are in
qualitative agreement with the experimental results on the spin
transitions \cite{park01,davenport12}. Within the model, the
ground-state spin transition occurs when the difference in the
Coulomb energies between the two ground states is balanced by the
change in the Zeeman energy due to spin polarization. The
calculations for $\nu=2/3$, 3/5, and 4/7 \cite{davenport12}
underestimate the critical Zeeman energy that determines the magnetic
field of the spin transition, $B^*$. In particular, these predict
$B^*\approx1.5$~T for $\nu=2/3$, assuming that the $g$ factor is
equal to $|g|=0.44$. The discrepancy between theory and experiment
cannot be explained by the finite thickness correction that is caused
by finite extent of the wave functions in the direction perpendicular
to the interface and gives rise to a decrease in the Coulomb energy
\cite{zhang86,morf02}.

In this paper, we employ a magnetocapacitance technique to determine
the magnetic field of the spin transition $B^*$ at filling factor
$\nu=2/3$. The ratio of the Zeeman and Coulomb energies is varied in
the experiment by a change of the back gate voltage that controls the
wave function extent through the steepness of the confining potential
in the direction perpendicular to the interface. In this experiment
one expects the Coulomb interaction effects to be revealed by their
contribution to the shift of the spin transition. We find that the
field $B^*$ decreases appreciably as the wave function extent is
increased. The calculated change of the Coulomb energy and $g$ factor
change because of nonparabolicity effects make approximately equal
contributions to the shift of the spin transition field. The two
mechanisms together describe the observed relative shift of $B^*$
with no fitting parameters. The case of the spin transition includes
less parameters and allows more rigorous comparison of experiment
with theory as compared to the case of fractional gaps. We check that
the $\nu=2/3$ gap at fields above $B^*$ increases with increasing
wave function extent and reaches the value of the $\nu=1/3$ gap,
indicating the gap symmetry in the high-field limit over the entire
range of back gate voltages studied.

\begin{figure}
\scalebox{0.34}{\includegraphics{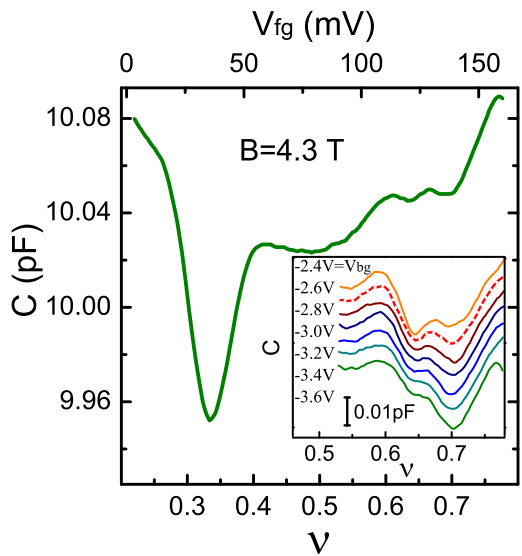}}
\caption{\label{fig1} Magnetocapacitance as a function of front gate
voltage converted into filling factor in $B=4.3$~T at $V_{bg}=-3.6$~V
and $T=60$~mK. The inset shows an expanded view on $C(\nu)$ near
$\nu=2/3$ at different back gate voltages. The dashed line
corresponds to the transition. Linear contribution to $C(\nu)$ is
subtracted and the traces are vertically shifted for clarity.}
\end{figure}

Measurements were made in an Oxford dilution refrigerator with a base
temperature of $\approx 30$~mK on remotely doped
GaAs/Al$_x$Ga$_{1-x}$As ($x=0.336$) single heterojunctions (with a
low-temperature mobility $\approx4\times 10^6$~cm$^2$/Vs at electron
density $9\times 10^{10}$~cm$^{-2}$) having the rectangular
geometry with area $1.8\times 10^4$~$\mu$m$^2$. The depth of the 2D
electron layer was 200~nm. A highly doped ($1\times
10^{18}$~cm$^{-3}$ Si) layer with thickness 100~nm was buried in the
bulk of GaAs, at a distance 5~$\mu$m from the interface. This layer
remained well-conducting at low temperatures and served as a back
electrode. A 200~nm low-temperature grown GaAs (LT-GaAs) layer
between the back gate and the 2D electron layer was used to block
leakage currents. The Fermi level in this layer is pinned near the
midgap, which results in the formation of a Schottky barrier between
$n$-GaAs and LT-GaAs \cite{maranowski95}. A metallic front gate was
deposited onto the surface of the sample. The presence of gates
allowed variation of both the electron density and the confining
potential by applying a dc bias between the gate and the 2D
electrons. The front gate voltage was modulated with a small ac
voltage of 2.5~mV at frequencies in the range 0.6--21~Hz, and both
the imaginary and real components of the current were measured with
high precision ($\sim10^{-16}$~A) using a current-voltage converter
and a lock-in amplifier. Smallness of the real current component as
well as proportionality of the imaginary current component to the
excitation frequency ensure that we reach the low-frequency limit and
the measured magnetocapacitance is not distorted by lateral transport
effects. A dip in the magnetocapacitance in the quantum Hall effect
is directly related to a jump of the chemical potential across a
corresponding gap in the spectrum of the 2D electron system
\cite{smith85}:
\begin{equation}\frac{1}{C}=\frac{1}{C_0}+\frac{1}{Ae^2dn_s/d\mu},\label{C}\end{equation}
where $C_0$ is the geometric capacitance between the gate and the 2D
electrons, $A$ is the sample area, and the derivative $dn_s/d\mu$ of
the electron density over the chemical potential is the thermodynamic
density of states. Near the filling factor $\nu=1/2$, the capacitance
$C$ in the range of magnetic fields studied reaches its high-field
value determined by the geometric capacitance $C_0$. The chemical
potential jump $\Delta\mu$ for electrons at fractional filling factor
is determined by integrating the magnetocapacitance over the dip (for
more details, see Ref.~\cite{khrapai07}):
\begin{equation}\Delta\mu=\frac{e}{C_0}\int_{\text{dip}}(C_0-C)dV_{fg}.\label{Delta}\end{equation}
Note that the excitation gap corresponds to the chemical potential
discontinuity divided by the fraction denominator \cite{halperin83}.

\begin{figure}
\scalebox{0.34}{\includegraphics{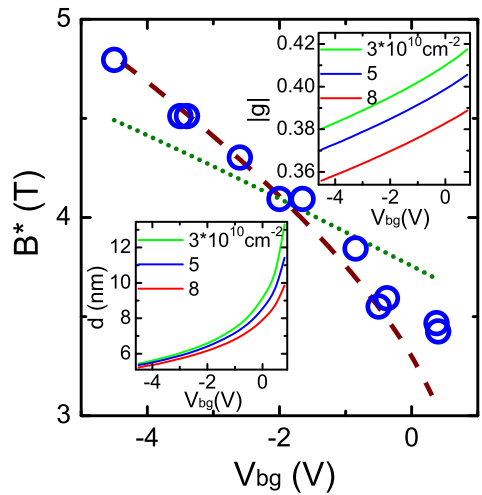}}
\caption{\label{fig2} Dependence of the magnetic field of the spin
transition at $\nu=2/3$ on back gate voltage. The symbol size
reflects the experimental uncertainty. The field $B^*$ versus
$V_{bg}$ is calculated using Eq.~(\ref{B*}) with $F=1$ (dotted green
line) and the factor $F$ taken from Ref.~\cite{davenport12} (dashed
brown line), where the coefficient $\alpha$ is chosen so that $B^*$
at $V_{bg}=-2$~V corresponds to its experimental value. The
calculated extent of the electron density distributions and
single-electron $g$ factor versus $V_{bg}$ are displayed in the
insets for different electron densities.}
\end{figure}

A magnetocapacitance trace $C$ as a function of front gate voltage
$V_{fg}$ converted into filling factor $\nu$ is displayed in
Fig.~\ref{fig1} for a magnetic field of 4.3~T and back gate voltage
$V_{bg}=-3.6$~V. A narrow dip and a double minimum in $C$ are seen at
$\nu=1/3$ and $\nu=2/3$, respectively. The double minimum feature in
$C(\nu)$ at different back gate voltages is shown in the inset to
Fig.~\ref{fig1}. At fixed electron density (or magnetic field), the
change of the confining potential is caused by a change of $V_{bg}$
that is connected to a shift of $V_{fg}$ via $\Delta V_{fg}\approx
-0.05\Delta V_{bg}$. One can see from the figure that the two minima
reveal an interplay with changing back gate voltage. The interplay
reflects the competition of the domains of two ground states at the
critical point \cite{chakraborty00}, similar to the case of the
integer quantum Hall effect \cite{piazza99,poortere00,khrapai00}.
This corresponds to the spin transition in the ground state
\cite{khrapai07}. We determine the transition point as a point at
which the two minima are approximately equal to each other (the
dashed line in the figure).

In Fig.~\ref{fig2}, we plot the field $B^*$ of the spin transition
for $\nu=2/3$ versus back gate voltage. The experimental value $B^*$
decreases approximately linearly with increasing back gate voltage
and remains well above the theoretical prediction $B^*\approx1.5$~T.
The increase in $V_{bg}$ decreases the steepness/slope of the
confining potential and increases the wave function extent. Based on
the competition between the Coulomb and Zeeman energies, one can
tentatively expect that the reduction in the Coulomb energy due to
the finite thickness of the 2D electron layer is responsible for the
observed shift of the spin transition. However, the absolute value of
$g$ factor should increase with increasing wave function extent,
which can contribute to the shift of $B^*$.

\begin{figure}
\scalebox{0.34}{\includegraphics{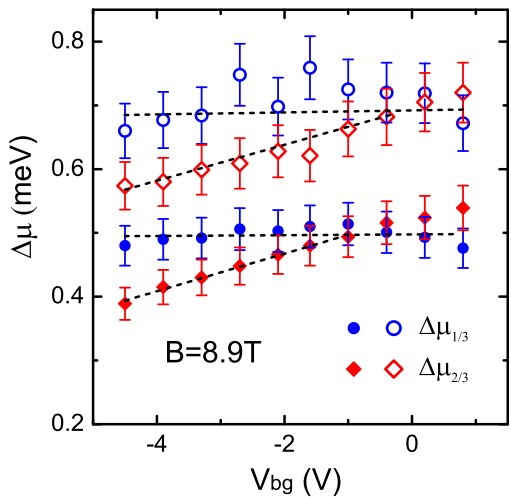}}
\caption{\label{fig3} The chemical potential jump at $\nu=1/3$ and
$\nu=2/3$ versus back gate voltage for $B=8.9$~T and $T=0.2$~K (solid
symbols). Also shown by the open symbols is the corrected data for
$\Delta\mu$, see text. The dashed lines are guides to the eye.}
\end{figure}

We verify the importance of spin effects by measuring the chemical
potential jump at $\nu=1/3$ and $\nu=2/3$ in magnetic fields above
$B^*$ as a function of back gate voltage, as shown by the solid
symbols in Fig.~\ref{fig3}. The data correspond to the limit of low
temperatures where $\Delta\mu$ saturates and becomes independent of
temperature \cite{khrapai07}. Whereas the gap at $\nu=1/3$ remains
approximately constant, the $\nu=2/3$ gap increases with back gate
voltage and reaches the value of the $\nu=1/3$ gap. The different
behavior of the gaps signals the presence of spin-dependent
contribution to the value of gap at $\nu=2/3$, i.e., the increase of
the $\nu=2/3$ gap with $V_{bg}$ should be related to the increase in
the absolute value of $g$ factor. We do not observe in the experiment
a decrease of the gaps with increasing back gate voltage, as expected
from the suppression of gaps due to the finite thickness correction.
This can be attributed to the manifestation of disorder effects: as
the 2D electrons are pushed closer to scatterers at the interface,
the gap suppression caused by disorder becomes stronger, compensating
roughly for the finite thickness correction. It is possible to take
into account the effect of long-range disorder potential which leads
to broadening the chemical potential jump as a function of filling
factor. In the spirit of Ref.~\cite{khrapai08}, linear extrapolations
of the dependence $\mu(\nu)$ to the fractional filling factor yield
$\Delta\mu$ for an ideal/homogeneous 2D electron system. The
corrected data for $\Delta\mu$ is shown by the open symbols in
Fig.~\ref{fig3}. One can see an upward shift in the dependences
$\Delta\mu(V_{bg})$ for both $\nu=1/3$ and $\nu=2/3$ so that their
behavior remains basically the same. Therefore, it is the effect of
short-range disorder potential that can be responsible for the
compensation of the finite thickness correction.

\begin{figure}
\scalebox{0.38}{\includegraphics{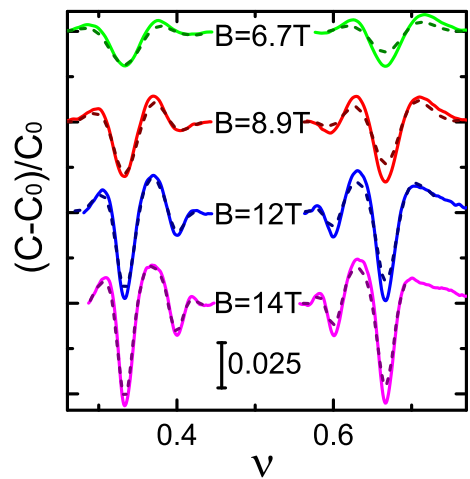}}
\caption{\label{fig4} Magnetocapacitance $(C-C_0)/C_0$ as a function
of filling factor for $V_{bg}=-4.5$~V (dashed lines) and 0.8~V (solid
lines) at different magnetic fields and temperatures 0.2, 0.2, 0.25,
and 0.3~K (top to bottom). The curves are vertically shifted for
clarity.}
\end{figure}

Figure~\ref{fig4} displays the magnetocapacitance $(C-C_0)/C_0$
versus $\nu$ measured in the low-temperature limit at $V_{bg}=-4.5$~V
(dashed lines) and 0.8~V (solid lines) in different magnetic fields.
As $B$ is increased, the minima at $\nu=1/3$ and $\nu=2/3$ become
symmetric and the value of the $\nu=2/3$ gap reaches that of the
$\nu=1/3$ gap. Thus, the gap symmetry is the case at the highest
magnetic fields over the entire range of back gate voltages studied.
This implies that the spin-dependent contribution to the $\nu=2/3$
gap changes for the Coulomb contribution.

We would like to emphasize the significant difference between the
theoretical descriptions of the ground-state spin transition and
fractional gaps. The experimental values of fractional gaps are known
to be noticeably smaller than the theoretical predictions, which can
be attributed to disorder effects. In contrast, the magnetic fields
of the spin transition measured in experiment exceed by far the
theoretical values, which cannot be explained by the disorder-caused
suppression of the interaction effects. Also, the $g$ factor related
to the spin transition is equal to its single-electron value, whereas
the gap-related $g$ factor can be enhanced due to interactions.
Therefore, the case of spin transition includes less parameters and
allows more rigorous comparison of experiment with theory.

It is easy to calculate the behavior of the single-electron $g$
factor with back gate voltage for the samples used. By solving the
Schrodinger and Poisson equations self-consistently \cite{snider90},
one obtains the dispersion, $d$, of the electron density distributions
perpendicular to the interface as a function of back gate voltage for
different $n_s$ (bottom inset to Fig.~\ref{fig2}). The value $d$
gives a measure of the wave function extent and corresponds to the 2D
subband bottom. This increases sharply with back gate voltage at high
$V_{bg}$, close to the voltage $V_{bg}\approx0.9$~V for our samples,
where the Schottky barrier between $n$-GaAs and LT-GaAs vanishes and
the gate leakage current arises. In our case the variation of the $g$
factor caused by wave function penetration in the AlGaAs barrier is
small, and the dominant $g$ factor change relative to $g=-0.44$ in
bulk GaAs originates from nonparabolicity effects. Within
the $kp$-theory, the change at the 2D subband bottom is equal to
$\Delta g=\langle T\rangle/W$, where $\langle T\rangle$ is the
average kinetic energy and $W$ is the characteristic energy
\cite{ivchenko92}. Using the experimental value of $W\approx150$~meV
\cite{dobers88,nefyodov11} and taking account of the
size-quantization energy and the diamagnetic shift by half the
cyclotron energy \cite{rem1}, we get the dependence of $g$ on back
gate voltage at different $n_s$, shown in the top inset of
Fig.~\ref{fig2}.

The behavior of the field $B^*$ with back gate voltage is determined
from the relation \cite{davenport12}
\begin{equation}\alpha F\frac{e^2}{4\pi\kappa_0\kappa l_B}=\frac{1}{2}g\mu_BB,\label{B*}\end{equation}
where the coefficient $\alpha$ is given by the difference in the
Coulomb energies of fully and partially polarized states and the
finite thickness correction $F=1$ ($F<1$) for zero (nonzero)
thickness of the 2D electron layer. Assuming that $F=1$ and choosing
the coefficient $\alpha$ so that $B^*$ at $V_{bg}=-2$~V corresponds
to its experimental value, we determine the relative shift of $B^*$
caused by the $g$ factor change (the dotted green line in
Fig.~\ref{fig2}) \cite{rem2}. One can see in the figure that this
mechanism accounts approximately for half of the observed shift of
the magnetic field of the spin transition.

We now take into account the suppression of the Coulomb interactions
due to the finite thickness of the 2D electron layer. Replacing the
calculated wave function by the Fang-Howard wave function
$\zeta(z)=(b^3/2)^{1/2}z\exp(-bz/2)$ (where the $z$-axis is
perpendicular to the interface, $z>0$ in substrate, and $b$ is a
parameter) \cite{ando82} with the same dispersion $d$, one determines
the value $1/bl_B=d/\sqrt3l_B$ that controls the suppression factor
$F$. Using the dependence $F(1/bl_B)$ calculated in
Ref.~\cite{davenport12}, we obtain from Eq.~(\ref{B*}) the total
relative shift of $B^*$ (the dashed brown line in Fig.~\ref{fig2}),
which is in agreement with the experiment \cite{remark}. Thus, both
mechanisms make approximately equal contributions to the change of
the spin transition field and describe the observed relative shift of
$B^*$ with no fitting parameters.

The theoretical magnetic field of the spin transition
\cite{davenport12} is approximately two times smaller than the value
observed in the experiment. That is to say, the theory in question
underestimates by $\approx30$\% the difference in the Coulomb
energies between fully and partially polarized states, which is
experimentally equal to $\approx0.005e^2/4\pi\kappa_0\kappa
l_B\approx0.04$~meV. The discrepancy with the experiment may result
from distant extrapolations used to determine the Coulomb energies.
Note that the afore-mentioned significant difference in the fields
$B^*$ for different samples
\cite{eisenstein90,clark90,engel92,kukushkin99,freytag01,groshaus07}
as well as the lack of symmetry between the $\nu=1/3$ and $\nu=2/3$
gaps in the same magnetic field (see, e.g.,
Ref.~\cite{chakraborty00}) are likely to be caused mainly by the
reduced absolute values of $g$ factor for the particular sample
design.

The gap-related $g$ factor can be determined assuming that the sum of
the gaps at $\nu=1/3$ and $\nu=2/3$ is equal to the Zeeman energy.
Using the corrected data in Fig.~\ref{fig3}, one estimates an
enhanced $g$ factor at $g\approx0.9$. Its variation $\delta
g\approx0.1$ in the range of back gate voltages studied is bigger
than the change of the single-electron $g$ factor (top inset of
Fig.~\ref{fig2}). This can be explained qualitatively by the
many-body enhancement of the $g$ factor, depending on the disorder
and finite thickness effects.

In summary, we have found that the magnetic field of the spin
transition at filling factor $\nu=2/3$ decreases appreciably as the
wave function extent controlled by back gate voltage is increased.
Our calculations show that the contributions to the shift of $B^*$
from the change of the Coulomb energy and the $g$ factor change
because of nonparabolicity are approximately the same. The observed
relative shift of $B^*$ is described with no fitting parameters. We
have checked that the $\nu=2/3$ gap at fields above $B^*$ increases
with increasing wave function extent and reaches the value of the
$\nu=1/3$ gap, indicating the gap symmetry in the high-field limit
over the entire range of back gate voltages studied.

We gratefully acknowledge discussions with S.~C. Davenport and V.~T.
Dolgopolov. We are grateful to J.~P. Kotthaus for an opportunity to
use the clean room facilities at LMU Munich. This work was supported
by RFBR 12-02-00272 and 13-02-00095, RAS, and the Russian Ministry of
Sciences.

\end{document}